# Role of Hydration on the Electronic Transport through Molecular Junctions on Silicon

*Nicolas Clément\*, David Guérin, Stéphane Pleutin, Sylvie Godey and Dominique Vuillaume\**

IEMN-CNRS, avenue Poincaré, Cité scientifique, Villeneuve d'Ascq, 59652, France

\*) nicolas.clement@iemn.univ-lille1.fr ; dominique.vuillaume@iemn.univ-lille1.fr

**ABSTRACT.** Molecular electronics is a fascinating area of research with the ability to tune device properties by a chemical tailoring of organic molecules. However, molecular electronics devices often suffer from dispersion and lack of reproducibility of their electrical performances. Here, we show that water molecules introduced during the fabrication process or coming from the environment can strongly modify the electrical transport properties of molecular junctions made on hydrogen-terminated silicon. We report an increase in conductance by up to three orders of magnitude, as well as an induced asymmetry in the current-voltage curves. These observations are correlated with a specific signature of the dielectric response of the monolayer at low frequency. In addition, a random telegraph signal is observed for these junctions with macroscopic area. Electrochemical charge transfer reaction between the semiconductor channel and $H^+/H_2$ redox couple is proposed as the underlying phenomenon. Annealing the samples at 150°C is an efficient way to suppress these water-related effects. This study paves the way to a better control of molecular devices and has potential implications when these monolayers are used as hydrophobic layers or incorporated in chemical sensors.

# I - INTRODUCTION

Nanofabrication facilities and newly discovered or developed nano materials offer large expectations for future applications in nanolectronics,[1] chemistry[2] or medicine.[3] However, parasitic effects that can substantially alter signals and electrical properties must be identified. Water molecules are one of them because of their presence in ambient atmosphere, and their strong affinity with many surfaces.[4-7] For instance, water and oxygen molecules are a key issue for organic electronics devices such as organic light emitting diodes or organic field-effect transistors, because organics are usually reacting with these molecules. To avoid degradation of the device performances, encapsulation and/or passivation processes are used and need to be further developed.[8] On the contrary, organic sensors must work in air and liquid environments and one needs to develop materials and devices which are compatible with the presence of water. At a molecular-scale, reports on the hydration effects on the electrical properties of molecular devices are scarce. Aguire et al. have reported that the oxygen/water redox couple can suppress electron conduction in carbon nanotube FET transistor (CNTFETs).[6] CNTFETs on $SiO_2$ were converted from unipolar to ambipolar by annealing the devices in a vacuum probe station for 24 h at 120 °C. The electrical characteristics were then measured either in high vacuum conditions or while exposing the devices to dry $O_2$, degassed $H_2O$ and dry $N_2$, at pressures corresponding to their partial pressures in air. These results have demonstrated that the $O_2/H_2O$ layer must be present at the oxide surface to induce electron conduction suppression in CNTFETs. This finding implies that electrons are not systematically trapped by hydroxyl groups, as originally proposed by Chua et al.[9]

To the best of our knowledge, only two groups have reported the effect of hydration on the transport properties of self-assembled monolayer molecular junctions using alkylthiols on gold[10] and Al/AlOx/ marcaptohexanoic acid/Au junctions.[11] They suggested that the main mechanism is a weakening of the interface bond, leading to a decrease of current in the junctions.

Here we show, on the opposite, that hydration induces a strong increase in the current for molecular junctions made of alkene monolayers grafted on hydrogenated silicon substrates. Since the pioneering paper by Lindford et al. on the grafting of organic monolayers on hydrogen-terminated, oxide-free, silicon surfaces,[12,13] the chemical and electrochemical functionalisation of these oxide-free Si surfaces have been widely studied to fabricate controlled organic/silicon interfaces of interest for various applications (see a review in Refs. [14,15]). Regarding electronics devices, Lewis and coworkers,[16,17] Sieval et al.,[18] reported that alkylation of H-Si surfaces allows fabricating low recombination velocity (<25 cm.s$^{-1}$, i.e. with a low surface trap density < 3x10$^9$ cm$^{-2}$), air-stable (good passivation against reoxidation) surfaces. Allongue and coworkers explored the properties of these organic monolayers as capacitors at the interface between silicon and aqueous electrolytes.[19-21] Many groups reported on their static (dc) current-voltage curves to determine their electronic structures at the interface with silicon and the electron transfer mechanism though them (e.g. tunneling vs. thermo-ionic, HOMO vs. LUMO mediated).[22-29] Here, we report that hydration can explain several behaviors reported for the dc and dynamic charge transport and electronic properties of such junctions. In particular, we show that hydration can explain the large dispersion of the dc current measured for monolayers having only minor changes of their structural quality.[30] We also demonstrate that hydration can be related to the appearance of a large interfacial relaxation peak observed in the low-frequency dielectric relaxation spectroscopy measurements.[31,32] Finally, we report the observation of an unusual Random-Telegraph-Signal (RTS) noise for macroscopic molecular junctions (lateral size of about few hundred micrometers), which cannot be explained by the conventional theory of single-electron trapping/detrapping[33] applied strictly for tiny devices (deeply submicronic). We propose that this anomalous RTS can be related with hydration effect, in close correlation with the other experimental behaviors mentioned *supra*. We show that the hydration can occur during the monolayer fabrication, even though we work in strictly anhydrous conditions. Thus, these features of the electronic properties of our molecular junctions are clear fingerprints of hydration. Annealing at 150°C is an efficient way to reduce the number of water molecules located in the molecular junction, leading to a suppression of all these electrical features. These effects are reversible since hydration with a droplet of water allows a complete recovery of these above mentioned fingerprints. In our junctions, we propose that a water-related energy level is created in the HOMO/LUMO[22] gap of the alkene molecules at Si interface, which we suggest to be the $H^+/H_2$ redox couple. We also found that hydration can be minimized using a low doped Si substrate and highly dense monolayer. In addition to improve the fundamental understanding of the role of water at the Si interface and of the electronic properties of molecular electronics on silicon, these results have a potential impact when such molecular monolayers are incorporated in chemical sensors.

## II - EXPERIMENTAL SECTION

Si-C linked alkyl monolayers were formed on n-type Si(111) substrates by thermally induced hydrosilylation of octadecene (18 carbon atoms chain, $C_{18}$) on hydrogenated silicon surface Si:H.[24,30] Three different samples were prepared: monolayers on low-doped substrates (1 $\Omega$.cm) immersed 30 mn (sample A) and 5 hours (sample B) in alkene solution, and monolayers on highly doped substrate (0.01-0.1 $\Omega$.cm) immersed 5 hours (sample C) in alkene solution. Prior electrical measurements, the monolayers were characterized by usual physicochemical techniques. Thickness measurements by ellipsometry and static water contact angle (C.A.) show that alkyl monolayers quality is comparable to similar monolayers reported in the literature (Table 1). For instance, according to detailed structural studies,[18] C.A. of 110° and thickness of 2 nm indicates a densely-packed monolayer. To investigate the changes on electrical properties due to hydration, all the samples were considered under three different conditions: "as fabricated" (reference), annealed at 150°C for one hour under dry nitrogen atmosphere (<5 ppm of $H_2O$) and hydrated with a droplet of deionized water (DI) for 20 s.

*Chemicals:* 1-tetradecene (99.8 %), 1-octadecene (99.5 %), hydrogen peroxide (30 % in water) were purchased from Sigma Aldrich. Alkenes were distilled over calcium hydride, dried over freshly activated 3 Å molecular sieves for at least 3 days and stored in a M-Braun glove box (<1 ppm $H_2O$ and $O_2$). Acetone (>99.8 %, VLSI grade), sulfuric acid (96 %, VLSI grade) were obtained from BASF.

Dichloromethane (99.8 %), chloroform (99 %), hexane (98.5 %), ammonium fluoride (40 %, electronic grade) were purchased from Carlo Erba.

*Silicon substrates cleaning:* Substrates were n-type Si(111) wafers purchased from Siltronix with resistivity of 0.01-0.1 $\Omega$.cm (high doping) and resistivity of 1 $\Omega$.cm (low doping). First, silicon substrates were sonicated in acetone, then chloroform for 5 min, then blown dry under nitrogen. Then, the samples were dipped into a freshly prepared piranha solution ($H_2SO_4$-$H_2O_2$ 2:1 v/v) at 100 °C for 1 h, rinsed thoroughly with DI water and etched in degassed 40 % $NH_4F$ solution for 10 min. This treatment (oxidation with piranha, rinsing, then etching with $NH_4F$) was repeated twice. Hydrogenated silicon substrates were perfectly dried under nitrogen flow and used immediately for hydrosilylation. *Caution: the piranha solution violently reacts with organic chemicals; consequently, it should be handled with extreme care.*

*Monolayers preparation:* Monolayers were prepared by thermal hydrosilylation reaction. These air-sensitive reactions were performed in a schlenk line (vacuum/nitrogen manifold). Schlenkware was perfectly dried in an air oven at 120°C overnight prior to use. The alkenes were degassed by 4-5 freeze-pump-thaw cycles. The freshly prepared Si-H substrate was immerged in neat deoxygenated alkene at 200°C (for octadecene) or 150°C (for tetradecene) during 30 mn for sample A and 5 h for samples B and C. Functionalized substrates were then removed, cleaned thoroughly by sonication for 3 min in hexane then dichloromethane and finally dried under nitrogen flow.

*Electrical measurements:* The hanging naked Hg drop (surface $S \sim 2.6 \times 10^{-4}$ cm²) was used as a metallic top electrode and was formed using a controlled growth mercury electrode system (BASI) in a glove box under $N_2$ flow ($H_2O$<50 ppm). The sample was placed on an elevator stage and put into contact with Hg by moving it upward. The junction size was estimated by measuring the Hg drop profile by a calibrated USB webcam. Current-voltage curves were acquired with an Agilent 4156C semiconductor parameter analyzer, with bias voltage swept from -1.5V to 1.5 V with a 10 mV step and integration time set to long. The capacitances were measured with an Agilent 4284A LCR-meter. For noise measurement the voltage *V* was applied with an ultralow-noise dc power supply (Shibasoku PA15A1). The source current was amplified with a DL 1211 current preamplifier supplied with batteries. Random telegraph signal (RTS) data and noise spectra were acquired with an Agilent 35670 dynamic signal analyzer. A detailed protocol for our noise measurements is shown in ref 34.

*Statistical analysis:* 1 sample (51 measures), 1 sample (75 measures) and 3 samples (94 measures) have been used for statistical analysis on samples A, B and C, respectively. A new Hg droplet was formed for each measurement to reduce eventual time-dependent effect of Hg oxidation on electrical properties. For the different measurements, the position of Hg droplet on the samples (~1 cm²) was moved randomly.

*Contact angle measurements:* Static water contact angle measurements were performed using a remote computer-controlled goniometer system (DIGIDROP by GBX, France). These measurements were carried out in a class ISO 6 clean room where humidity, temperature, and pressure were controlled. Static drops of deionized water (18 MΩ.cm, 1-10 µL) were applied to the functionalized surfaces with a micropipette and the projected image was recorded by the remote computer. Contact angles were then extracted by contrast contour image analysis software. These angles were determined 5 s after application of the water drop. Measurements made across the functionalized surfaces were within 1°.

*Spectroscopic ellipsometry:* Spectroscopic ellipsometry data in the visible range was obtained using a UVISEL by Jobin Yvon Horiba Spectroscopic Ellipsometer equipped with DeltaPsi 2 data analysis software. The system acquired a spectrum ranging from 2 to 4.5 eV (corresponding to 300-750 nm) with 0.05 eV (or 7.5 nm) intervals. Data were taken using an angle of incidence of 70°, and the compensator was set at 45.0°. To determine the monolayer thickness, we used the optical properties of silicon from the software library, and for the organic monolayer we used the refractive index of 1.50. Accuracy of the monolayer thickness measurements is estimated to be ± 1 Å.

*XPS measurements:* X-ray photoemission spectroscopy (XPS) measurements have been performed using a Physical Electronics 5600 spectrometer to check the absence of oxidized silicon or any unremoved contaminants (the pressure in the analysis chamber was $6.10^{-10}$ Torr). We used a monochromatic AlKα X-ray source ($h\nu=1486.6\ eV$) and an analyzer pass energy of 12 eV. The acceptance angle of the analyzer has been set to 14°, the detection angle was 45° and the analyzed area was defined by an entrance slit of 400 $\mu$m. Under these conditions, the overall resolution as measured from the FWHM of Ag3d5/2 core level (CL) is 0.55 eV. The CLs intensities were measured as the peak areas after standard background subtraction according to Shirley procedure. Peaks have been decomposed in Voigt functions by keeping constant Gaussian and Lorentzian broadenings and using a least-square minimization procedure.

## III - RESULTS

### III - 1. Signatures of hydration on electrical measurements

To measure the electrical properties of the Si/alkene junctions, we used a Hg drop top electrode as schematically shown in Fig. 1-a.[24,35,36] We performed more than 50 measurements per sample type (see experimental section for statistical analysis). The top electrode (Hg drop, see experimental section) is biased with voltage $V$ keeping Si grounded. We define a representative curve for each of our sample type as one of the curve belonging to the maximum of counts of the current histograms (all the curves are available in supporting information, Fig. S1 and histograms are shown in Fig.3). Representative current density-voltage $J$-$V$ curves for as fabricated A and B samples are shown in Fig. 1-b. The shape and current density variation between "as fabricated" samples (two orders of magnitude larger for A than for B at positive bias) are in agreement with previously reported results for alkene monolayers on hydrogenated Si made with the same protocol.[30] Note that, especially for sample A, we observe a clear change of the slope of the $J$-$V$ curve at around 0.4-0.5V, which has been ascribed to a change from a thermoionic emission at the Si/monolayer interface (at low bias) to a tunneling transport regime (at higher bias) through the monolayer in these molecular junctions.[24] Rectification, i.e. lower current at negative bias is related to the depletion layer in the low-doped silicon substrate.[37] Indeed, in that case, the current is no longer determined by the electron transfer properties of the monolayer (e.g. thermoionic emission and tunneling effect as mentioned above), but by diffusion in the silicon space charge layer. Note that small differences of the structural properties (a difference of 2° for the C.A. and of 0.3 nm for the thickness between samples A and B - see Table 1) seem inducing large variations of the current. A large current variation for apparent minor difference in monolayer quality/thickness was initially attributed to a large sensitivity of thermoionic emission mechanism.[30]

Another interesting experimental result is the capacitance-voltage (*C-V*) measurements shown in Fig. 1-c (representative curves for as fabricated devices, all the curves are shown in supplementary information, Fig. S1). Capacitance in accumulation (*V>0*) at 1 kHz for "as fabricated" sample A is increased by more than one order of magnitude compared to sample B. An increase of the low-frequency (*f* < 1 kHz) capacitance was already observed for deliberately bad structural quality alkylsilane monolayers on naturally oxidized silicon.[31] This feature has been explained by the possible formation of metallic nano-filaments though structural defects in the monolayer during the subsequent top metal electrode evaporation. Since we use Hg drop contact, this explanation can be ruled out. We recently studied interfacial relaxation dynamics in the same kind of molecular junctions using admittance spectroscopy measurements,[32] and we have observed that a large density of defects (about $2-3 \times 10^{11}$ cm$^{-2}$) at the alkene monolayer/silicon substrate is responsible for a 10 fold increase in the low-frequency (< 1 kHz) capacitance.

The third surprising experimental result is the observation of unusual random-telegraph-signal (RTS) noise, which we observe here for macroscopic molecular junctions (contact area of $3.6 \times 10^{-3}$ cm$^2$) (Fig.1-d). RTS is usually attributed to the trapping/detrapping of a single (or a few) electrons by a single (or a few) traps, and it is observed for sub-micrometric devices.[33] In large devices, a large number of RTS noise give the so-called low-frequency 1/f noise.[33] In the following, we will show that these anomalous experimental features can be related to hydration.

***Current - capacitance correlation plots.*** Fig. 2-a shows the current (at 1V)-capacitance (at 1 kHz) log-log plot for molecular junctions on sample A under the three different conditions defined above: "as fabricated", annealed and hydrated. Annealing/hydration cycle is repeated twice. Although we note a large data dispersion (especially for the reference and hydrated samples, see also histograms of current in Fig. 3), we can clearly observe two distinct populations, where the "as fabricated" and hydrated junctions have current densities and capacitances larger by about 2.5 and 1 orders of magnitude (respectively) than that of the annealed samples. "As fabricated" sample behave as the hydrated ones. Such hydration effect is not observed on sample B, as shown in Fig. 2-b (*I-V* and *C-V* curves are shown in supplementary information, Fig.S1). In that case, all data are concentrated at about the same *log(J)-log(C)* coordinate as the annealed sample A. *J-V* curves selected with a similar current level at 1 V for sample A (annealed) and sample B ("as fabricated") are almost superimposed. Therefore, the large difference of the electrical properties observed for "as fabricated" samples A and B, which only differ from minor changes in their monolayer quality, can be mainly related to hydration.

***Current statistics.*** Histograms of current are used for a more detailed study on the role of hydration on electrical characteristics. Histograms of current *log(J)* at +/- 1 V for sample A are shown in Fig. 3-a. Whereas the current for the hydrated samples is about two orders of magnitude larger than for the annealed ones for the majority of devices at +1 V, such effect is not clearly observed at – 1 V. The lower current at -1 V, with respect to the one at + 1V (rectification effect) is related to the depletion of carriers in the silicon at negative bias and are therefore only weakly related to the monolayer properties. To avoid the silicon depletion effect, we performed similar experiment on the monolayer grafted on highly doped substrate (sample C). Representative *log(J)-V* curves for hydrated and annealed samples are shown in Fig. 4-a (all curves are shown in supplementary information, Fig. S2). Histograms at -1 V and +1 V are shown in Fig. 4-b. For these samples, the electron transport for both bias polarities is controlled by the monolayer (mainly tunneling effect in the case of molecular junctions made on highly-doped Si[38]). For "as fabricated" samples, we observe two peaks in current histograms, with 3 orders of magnitude difference, that we attribute to junctions with and without presence of water. The highest peak is in the low current state after annealing and it is in the high current state after hydration. However annealing/hydration effect is not as clear as for sample A. It becomes clearer when reducing alkyl chain length. A complete set of data for molecular junctions on the same highly-doped Si substrate as sample C, but with $C_{14}H_{29}$ alkene molecules is shown in supplementary information, Fig. S3. Small differences compared to $C_{18}H_{37}$ (C18) molecules are found: whereas the effect of annealing is not efficient for all measured samples, all hydrated samples show the increase of their low-frequency capacitance and current. Since molecular packing is increased with the number of carbon atoms in the alkyl chains,[39] this result is consistent with the fact that C18 monolayer provides statistically a better barrier for water. As shown in ref.32, alkylsilane junctions made on silicon substrate with a very thin $SiO_2$ layer (6Å) have also the signature of hydration (increase of the low-frequency capacitance), although not identified as so at that time. This capacitance increase was removed with annealing in similar conditions as in the present study. Note that a small asymmetry is still observed for hydrated junctions on highly doped substrate (Fig.4-c). The rectification ratio *I(1V)/I(-1V)* is 14 for the hydrated samples and 1.5 for annealed ones (obtained from average on 8 and 9 samples, respectively).

**III - 2. Admittance spectroscopy**

To gain insights in the impact of water molecules on the electrical properties of the molecular junctions, we compare the admittance spectroscopy for hydrated and annealed samples A. Admittance spectroscopy is a powerful technique to assess the dynamic properties of organic monolayers, e.g. dipole relaxation in the monolayer and/or at the monolayer/Si interfaces.[32] The capacitance and conductance measured at 1 V versus the frequency are shown in Fig. 5-a and 5-b, respectively. The slight decrease of capacitance with frequency for the annealed monolayers was already investigated and attributed to molecular dipolar relaxation.[32,40] The large increase in capacitance at low frequency (< 10 kHz) has been attributed to a large increase in monolayer/Si interface states.[32] Here, we propose that these states can be water-related (see discussion section *infra*). This contribution is taken into account with a capacitance $C_{H2O}$ and a dynamic resistance $R_{H2O} = \tau/C_{H2O}$, where $\tau$ is the time constant related to the process discussed below. A simple equivalent circuit considering both contributions, adapted from ref. 32, is shown in Fig. 5-a, inset. The following equivalent admittance $Y_{TOT}$ is obtained:

$$Y_{TOT} = G_{SAM} + G_{H2O} + \frac{\omega^2 \tau C_{H2O}}{(1+\omega^2\tau^2)} + C_{SAM} tg\delta\omega + j\left[C_{SAM} + \frac{C_{H2O}}{(1+\omega^2\tau^2)}\right]\omega = G(\omega) + jC(\omega)\omega \quad (1)$$

where $\omega=2\pi f$ is the pulsation, $f$ the frequency, $C_{SAM}$ and $G_{SAM}$ are the intrinsic capacitance and conductance of the monolayers, $C_{H2O}$ and $G_{H2O}$ are the water-related capacitance and conductance, and $tg\delta$ the dielectric loss of the monolayer. Reasonable fits are obtained from capacitance and conductance with the parameters shown in Table 2. The large value obtained for $G_{H2O}$ compared to $G_{SAM}$ will be discussed below. Note that the weak slope of the *C-f* curve for the annealed sample is not well fitted (Fig. 5-a) because this behavior is mainly due to the molecular dipole relaxation that should be fitted with a more complex function.[40] We simply use the average value to fix the intrinsic capacitance of the monolayer ($C_{SAM}$). Finally at a so-called "corner frequency" $f_c \approx 1/2\pi\tau$ (here about 1 kHz), a step $\Delta G$ in $G(\omega)$ curve of the hydrated device starts to occur (inset of Fig.4-b), with an amplitude given by $\Delta G = C_{H2O}/\tau$. Note that $f_c$ also corresponds to the intercept of the asymptotic behaviors of the *C-f* curve (Fig. 5-a).

**III - 3. Random Telegraph Signal**

Two-level fluctuations of the current, the so-called Random-Telegraph-Signal (RTS), are observed on these molecular junctions either with Hg or evaporated Al upper gate (Fig. 6), which are both macroscopic electrodes (in the range $10^{-4}$ - $10^{-3}$ cm$^2$). Since Hg is known to prevent from eventual metallic shortage, the breaking/reformation of metallic nano-filaments in the monolayer cannot be at the origin of the observed RTS. In addition, RTS is not observed for annealed molecular junctions (see supporting information, Fig. S4). Usually, RTS is related to the trapping/detrapping of electrons by a single (or a few) traps and is only observed for tiny devices (< 1 $\mu m^2$), mainly in low temperature experiments.[33] Therefore, this unusual RTS evidences the necessity to consider an alternative origin, which is discussed in section IV.

**III - 4. XPS measurements**

Whereas only minor differences are observed with contact angle and ellipsometry thickness measurements (see Table 1), we performed XPS analysis to gain insight into the effects of hydration on the physico-chemical properties of the molecular junctions and establish relationship with electronic properties. XPS Si2p core level (CL) spectra acquired just after grafting the monolayer (Fig. 7-a) evidence a partial oxidation of the substrate for samples C associated to a very little signal in the 102-104 eV region corresponding to $SiO_x$ species, whereas Si-O bond signal is negligible for samples A and B. Locally oxidized areas (i.e. Si-OH silanol groups) are generated by reaction of ungrafted Si-H functions with water or by the combined reactions of oxygen and UV light when the samples are exposed to ambient air conditions. The chemical mechanisms of silicon oxidation were detailed by Wayner and Wolkov[41] and depend on the doping level of the silicon substrate[42]. The oxidation is larger for high doping level. This explains the difference of oxidation between samples A,B and C, where the only difference is the substrate doping level (A,B: low doped, C: highly doped). In addition, due to the strong Si-OH-$H_2O$ interaction, water molecules may be more difficult to remove for sample C under annealing or vacuum, which is in agreement with the observed partial effect of annealing (see statistics: Fig. 4-b) compared to the more complete effect observed for samples A (Fig. 3-a). The higher concentration of oxygen atoms in sample C is also confirmed by its higher [O1s]/[Si2p+C1s] ratio of 0.091 compared to 0.009 ratio for sample B (Table 1). Moreover, XPS measurements evidence differences in the chemical environment of oxygen atoms. Indeed, the O1s peak decompositions[43] (Fig.

7-b and Fig. 7-c) reveal that the oxygen signal is due to both -OH groups (532.1±0.05 eV) and $H_2O$ molecules (533.3±0.005 eV) adsorbed on the surface or in the monolayer. The $H_2O$/[Si2p+C1s] ratio (Table 1), indicates that there is more water for sample C than for sample A, sample B being an intermediate case.

**IV. PROPOSED MECHANISM AND DISCUSSION**

The detailed microscopic mechanism at the origin of the observed electronic properties may however be complex. Recent studies of the influence of physisorbed water on the conductivity of hydrogen-terminated silicon-on-insulator surfaces have mentioned that there is actually no consensus about the microscopic mechanism.[4,5] For instance, Hall measurements have pointed out that hydration mainly affects the carrier density in the silicon films (and not the carrier mobility), and three mechanisms have been suggested: i) the increase of surface carrier density arises from charged surface states inducing silicon band-bending ii) from an electrochemical charge transfer process iii) from the dipole moment of the water molecules. According to Fig. 1-b, the effect of charged surface states on silicon band-bending is only weak because we did not observe any significant voltage shift of the capacitance-voltage curves between the annealed and hydrated samples (see supplementary information, Fig. S1 for statistics on *C-V* curves). The small change in the silicon band-bending is also consistent with the small variation of the current at negative bias for the low-doped Si samples (A and B, Fig. 3-a and Fig. S1). Any variation of the silicon band-bending, and consequently of the width of the silicon space charge region would have induced a large variation of the current in the silicon depletion regime.

Here we propose a water-related mechanism, which is consistent with our results. We suggest that an energy level related to electrochemical charge transfer reaction between water-related species and electrons coming from the silicon electrode is introduced at the monolayer/Si interface. Figs. 8-a and 8-b show the *C-f* curves measured at various *V* and the *C-V* curve at 1 kHz for the hydrated sample C, respectively. A maximum of capacitance at 1 kHz is obtained at 0.9 V. This maximum in the capacitance may be explained if we assume that, at this voltage, the silicon Fermi level is in resonance[7] with an additional energy level $E_T$ induced by the presence of water at the monolayer/silicon interface (i.e. both energy levels are aligned so that electrons can tunnel back and forth between the conduction band and $E_T$ - Fig. 8-c, and this electron dynamics is detected by capacitance measurement). Such a maximum is not observed for the annealed samples (flat behavior, Fig. 8-b, as expected for a highly doped silicon substrate). Interestingly, from data shown in Fig. 8-a, the voltage dependence of the capacitance corner frequency ($f_c = 1/2\pi\tau$), shows a change of its slope at a similar voltage (Fig. 8-b). We can estimate the energy position of this water-related level $E_T$ by simply considering the energy band diagram shown in Fig. 8-c. According to Ref. 42, we consider a thickness of 2.5 Å for the first $H_2O$ monolayer linked to Si-OH.[43] The tunneling distance $y_T$ between electrons in silicon and the water-related level $E_T$ should be $0 < y_T < 2.5$ Å (Figs.8-c and d). For the sake of simplicity, we consider $y_T$ as half this thickness (i.e. the energy level $E_T$ centered at the middle of the water molecule). If we assume a constant electric field through the monolayer, $F_M$ (Fig. 8-c), we can estimate the position of the related energy level $E_T$ (related to the Fermi level in bulk silicon) writing:[44]

$$F_M = \frac{V_{RES} - \psi_S}{t_M} = \frac{E_T}{q \cdot y_T} \quad (2)$$

where $t_M$ is the monolayer thickness, $V_{RES}$ is the voltage at resonance and $\psi_S$ is the surface potential in silicon at the same bias and $q$ is the elementary charge. Considering an explicit model for the voltage dependence of $\psi_S$ in accumulation regime[44] (see supporting information, Fig. S5), we have, at $V_{RES} = 0.9 \pm 0.15$ V, $\psi_S = 80 \pm 20$ mV. Taking $t_M = 19 \pm 1$ Å (see Table 1), $y_T = 1.25$ Å, , we get $39 < E_T < 56$ meV. If we refer this energy level to the vacuum level, we have to substract the electron affinity of the silicon, $\phi_{Si} = 4.05$ eV, and the energy difference between the conduction band and the Fermi level in the bulk Si, $\varepsilon_C - \varepsilon_F = kT.ln(N_C/N_D)$. With $N_C = 2.8 \times 10^{19}$ $cm^{-3}$ the equivalent density of electronic states in the conduction band of silicon (at 300K) and $N_D \approx 10^{18}$ $cm^{-3}$ the doping level of silicon (sample C), we get $E_{Tvac} \approx -4.09$ eV. This energy level is consistent with the energy level of $H^+/H_2$ redox couple located between -4.44 eV and -4 eV (Fig. 8-e) below vacuum level[45,46] depending on the *pH* of water, while the energy levels for the $H_2O/O_2$ redox couple are between -4.83 and -5.66 eV, not consistent with our experiments. These values do not consider interaction of water with the alkyl chains in the monolayer which may shift energy levels. The presence of such a resonant energy level, associated with the large dipoles of water molecules is also consistent with the large $G_{H2O}$ conductance compared to $G_{SAM}$, the large capacitance $C_{H2O}$ (see Table 2) and the asymmetry in *I-V* curves for hydrated samples, because this resonant energy level can induce a trap-assisted tunneling and tunnel barrier lowering which may be at the origin of such effects as already observed.[47-49] For sample A, $V_{RES}$ is shifted to higher bias ($V_{RES} \approx 1.35$ V $\pm 0.25$: see Fig. S6). It is due to the lower doping level of silicon that induces a shift of $V_{FB}$ to 0.4 V (Fig. 1) and a larger surface potential $\psi_S \approx 220 \pm 25$ mV at $V_{RES}$ (see supporting information, Fig. S5). Using Eq. (2), we get $E_T \approx 74$ meV and $E_{Tvac} \approx -4.19$ eV, again in the range of expectations for the energy level of $H^+/H_2$ redox couple.

The effect of hydration on the transport properties of molecular junctions has been recently reported for alkylthiol monolayers on gold surface.[10] These authors showed that the current decreased after hydration, opposite to what we have observed in our junctions on silicon. They suggested that the primary interaction affecting molecular conduction in akylthiol junctions is a rapid hydration at the gold-sulphur contacts, leading to a weakening of the Au-S bond and a decrease of charge transport. In our junctions, the molecules are covalently bonded to the silicon substrate. Therefore, this effect is unlikely. Due to a larger work function of the gold electrode compared to the electronic affinity of silicon, the above mentioned redox couple is far from accessibility for gold subtrates, which, as a consequence, leads to a different impact of hydration for alkylthiol molecules. We have checked that additional interfacial capacitance at low frequency is not observed for hydrated molecular junctions of alkylthiol on gold surfaces (see supporting information, Fig. S6).

We also propose that this water-related redox level can explain the unusual RTS noise in our molecular junctions. Usually, RTS is related to the trapping/detrapping of electrons by a single (or a few) traps in an insulting barrier at tunneling distance from an electron reservoir such as a transistor channel. The number of active traps has been derived by Mc Whorter[50] as $k.T.N_T(E).S/\gamma$ where $N_T$ [cm$^{-3}$.eV$^{-1}$] is the density of traps per unit volume and unit energy, $S$ is the channel area, $\gamma$ is the Mc Whorter tunneling constant (typically 1 Å$^{-1}$), $k$ the Boltzman constant and $T$ the temperature. RTS is only observed when there are only few active traps, either for tiny devices ($S < 1$ $\mu$m$^2$) or for devices with an extremely low trap density and mainly at low temperature (see a review in Ref. 33). The most studied case has been the silicon MOSFET,[33] where traps are typically dangling bonds in oxide. Each trapping-detrapping event induces a shift of gate voltage $|\Delta V|=\alpha.q/C$, where $q$ the electron charge, $C$ the device capacitance and $\alpha$ is a correction parameter described as $\alpha \approx \Delta S/S$ (typically <<1) with $\Delta S$ the channel area screening the trapped charge.[33] Recently, using an extremely small undoped nanowire transistor ($S \approx$ 15 nm x 50 nm) the limit $\Delta S=S$ has been reached and a correcting term has been introduced for $\alpha$ to account for dielectric constants.[34] In any case, for FET transistors, $\alpha <1$, except some exceptions for long 1D devices such as nanowires[51] or carbon nanotubes.[52] In the case of a MOS tunnel capacitor (a device similar to our molecular junctions with an ultra-thin silicon oxide instead an organic monolayer as the tunnel barrier),[53] the shift of voltage across the junction can be magnified ($\alpha$ up to 1000) due to image-charge induced tunnel barrier lowering[49,53] or trap-assisted tunneling.[54] Following Ref. 53, we have $|\Delta V|\approx\Delta I/(\delta I/\delta V)q/C$ where $\Delta I$ is the discrete variation of the current between the two levels (Fig. 8-a), $(\delta I/\delta V)$ the derivative of I-V curve at a given V. Here, we get $\alpha\approx 10^7$. This value is much larger than the 10$^3$ value usually observed for RTS in inorganic tunnel barriers.[49,53]

RTS has also been observed in single molecular junctions (STM).[55-58] In these experiments, the current fluctuations were ascribed to stochastic modifications of the conformation of the S-Au link at the molecule/Au interface.[55-57] Recently, STM experiments showed that these two-level fluctuations are not observed for molecule linked on hydrogenated Si through a Si-C bond,[58] mainly because the binding energy is larger for Si-C than for Au-S bonds.[59] Thus, we believe that we can discard this mechanism in our monolayer-based junctions. A more detailed analysis of the RTS noise in our molecular junctions allow us to correlate this RTS with the other fingerprints of the $H^+/H_2$ redox energy level, as observed from the capacitance-frequency $C$-$f$ measurements. As expected for two-level fluctuations for uncorrelated events, the upper and lower characteristic times ($\tau_{up}$ and $\tau_{down}$, see Fig. 6-a) follow a poissonian distribution (Fig. 6-b), with average values $\tau_{up} \approx 1$ ms and $\tau_{down} \approx 2.1$ ms. This RTS leads to a Lorentzian spectrum in the frequency domain (flat noise at low frequency and $1/f^2$ noise over the corner frequency $f_{cRTS}$) superimposed to the $1/f$ noise.[33] In Fig. 6-c, we show such behavior around the resonance voltage $V_{RES}$. If we compare the frequency behavior of the noise power spectrum with the $C$-$f$ behavior (Fig. 6-c), remarkably, both types of corner frequencies are quite similar ($f_{cRTS} \approx f_C$), which presumes of the same origin. Thus we can conclude that RTS is probably due to the presence of $H^+/H_2$ redox couple.

## V. - CONCLUSION

To conclude, we have shown experimentally that hydration has an important impact on the electronic transport of molecular junctions on silicon with an increase of the current (up to three orders of magnitude) and asymmetry in $I$-$V$ curves (ratio up to 14 at $|V|$=1 V) even in an atmosphere with less than 50 ppm of water. Since molecules are covalently bonded to the substrate, weakening of interface bond is unlikely. A water-induced energy level at the monolayer/Si interface is extracted from relaxation dynamics experiments. The measured value of this energy level is in good agreement with the known data for the $H^+/H_2$ redox couple. In addition, a Random Telegraph Signal, also attributed to this effect, is observed for molecular junctions with macroscopic electrodes. Since annealing at 150°C is an efficient way to remove this effect, this study paves the way to better controlled molecular devices and chemical sensors using such molecular monolayers. Large frequency dependence of capacitance is a direct way to detect the presence of water molecules in alkenes or other organic monolayers on silicon.

SUPPORTING INFORMATION AVAILABLE.

*J-V* curves for samples A, B and C, *C-V* curves for samples A,B, RTS for annealed samples, estimation of $\psi_s$, *C-f* for alkyl-thiol with GaIn upper electrode and data for molecular junctions with C14 molecules. This material is available free of charge via the internet at http://pubs.acs.org.

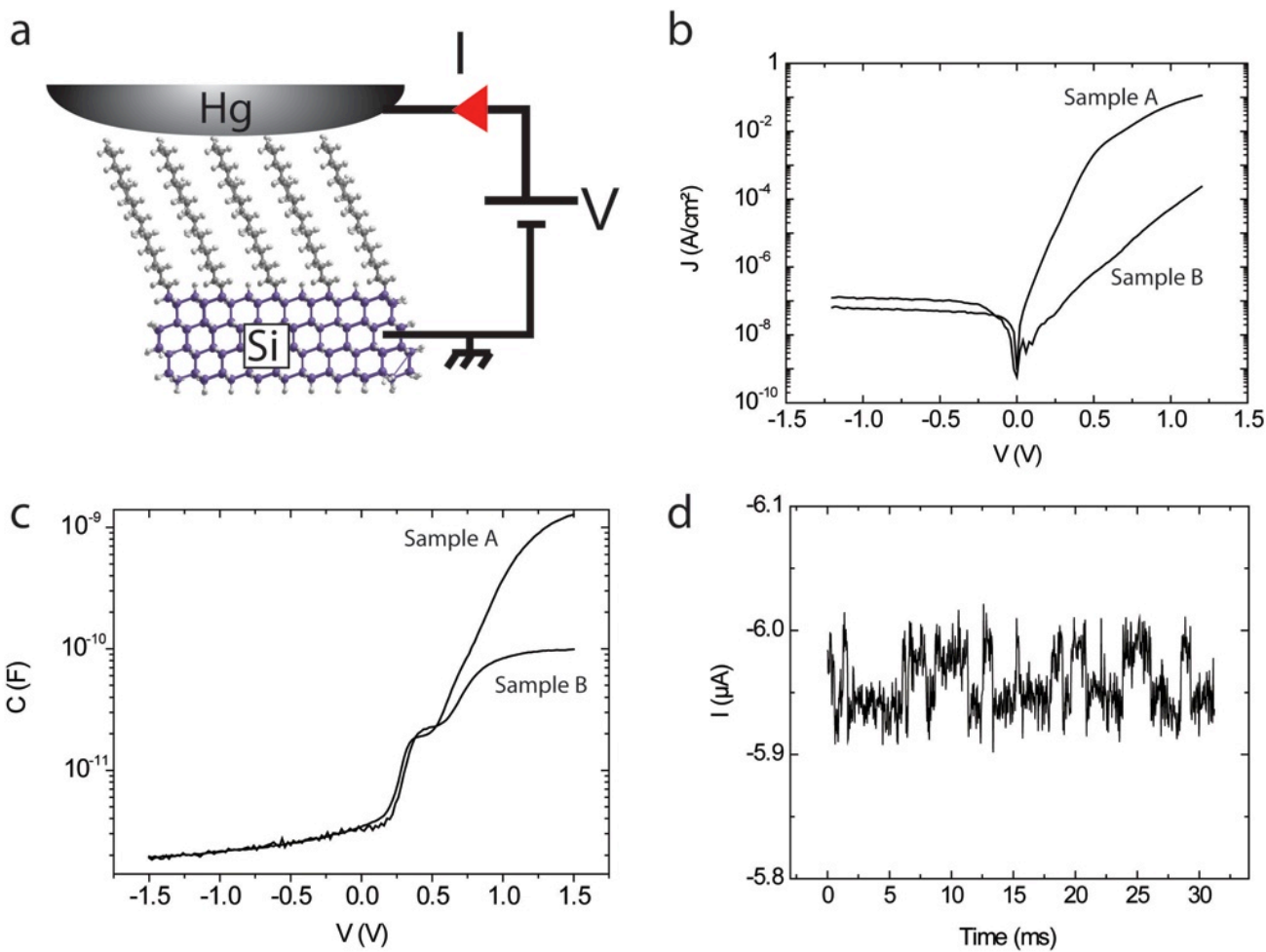

**Figure 1.** a) Schematic view of the molecular junction composed of silicon substrate, alkene monolayer and Hg drop. Si is grounded and a voltage *V* is applied to Hg drop contact. Experiments are performed in a glove box filled with $N_2$ (<50 ppm of $H_2O$).

b) *J-V* (current density-voltage) characteristics of freshly prepared ("as fabricated") samples A and B (representative curves).

c) *C-V* (capacitance at 1 kHz-voltage) characteristics of freshly prepared samples A and B (representative curves).

d) Random-Telegraph-Signal (RTS) observed for sample C measured at V= 0.9 V

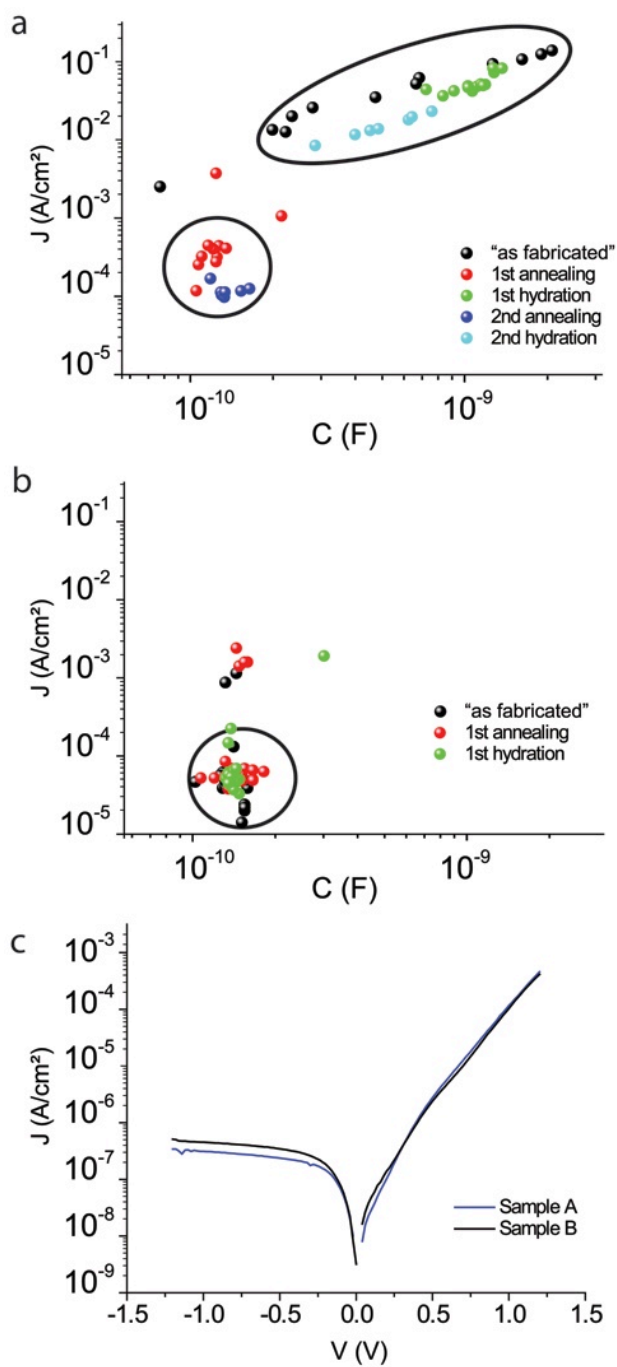

**Figure 2.** a) Current density - capacitance correlation plot, *log(J) (at 1 V) - log(C) (at 1kHz)* for two cycles of annealing/hydration (sample A)

b) *log (J) (at 1 V) - log(C) (at 1kHz)* correlation plot for two cycles of annealing/hydration (sample B)

c) *J-V* curves of a sample A (annealed) and a sample B (reference) selected for having the same *J* at 1 V

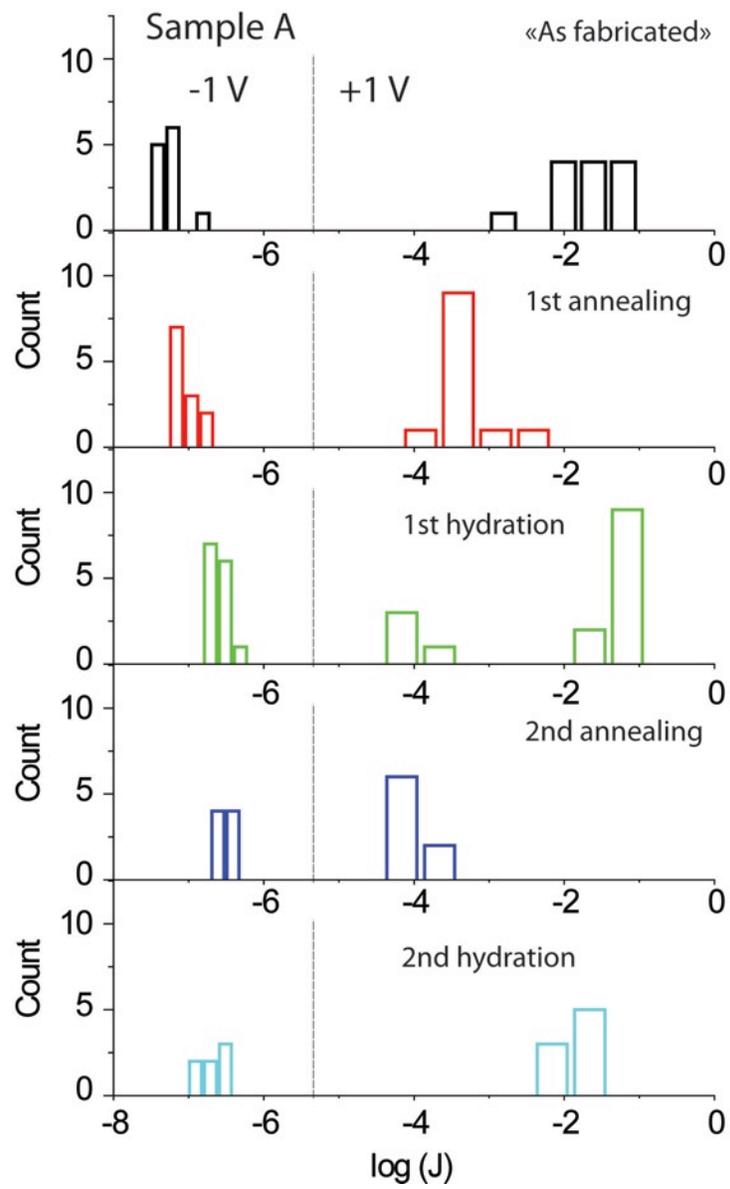

**Figure 3.** Histograms of log(*J*) for sample A including the 2 annealing/hydration cycles. The current density is measured at -1 V (left part) and + 1 V (right part).

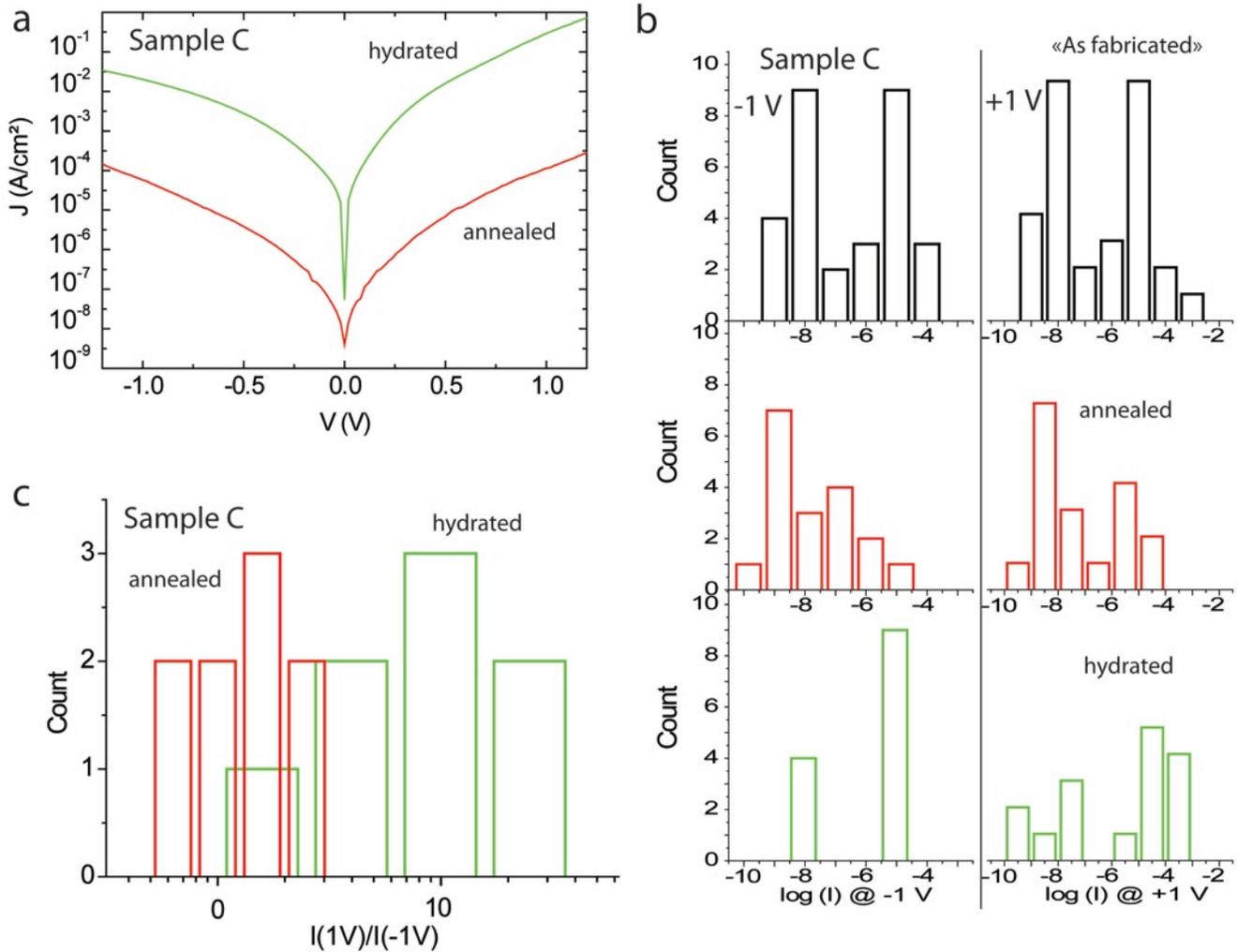

**Figure 4.** a) Representative current-density - voltage, *J-V*, curves for sample C (hydrated and annealed).

b) Histograms of log(I) for sample C (reference/annealing/hydration) at -1 V (left part) and +1 V (right part) for the "as fabricated" sample, after annealing and after hydration.

c) Histograms of the rectification ratio *I(1 V)/I(-1 V)* for sample C after annealing and a subsequent hydration.

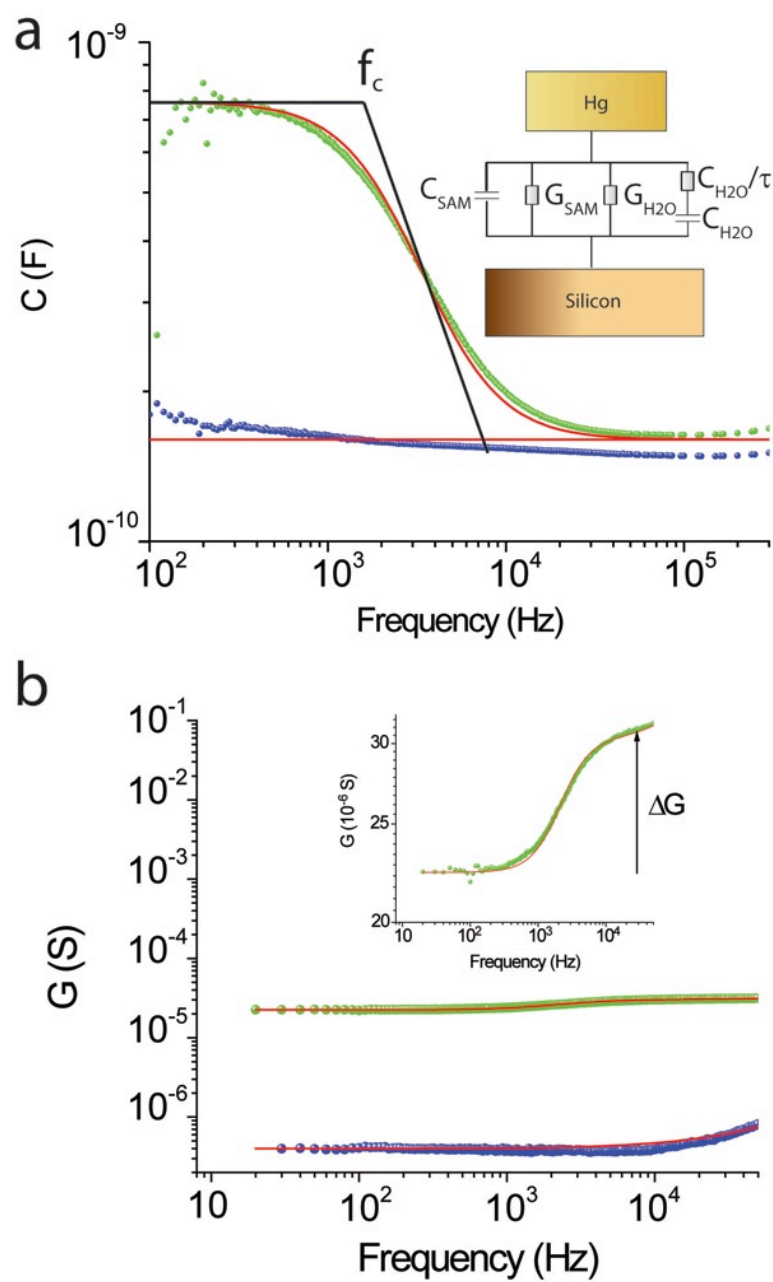

**Figure 5.** a) Capacitance-frequency and b) conductance-frequency measurements for annealed (blue dots) and hydrated (green dots) sample A. Red lines are the fits of the model given by Eq. 1 with parameters shown in table 2. Inset of a): Equivalent circuit model including the intrinsic monolayer

contribution and the water related contribution. Inset of b): zoom showing the step-like variation of conductance $\Delta G$ above a few kHz.

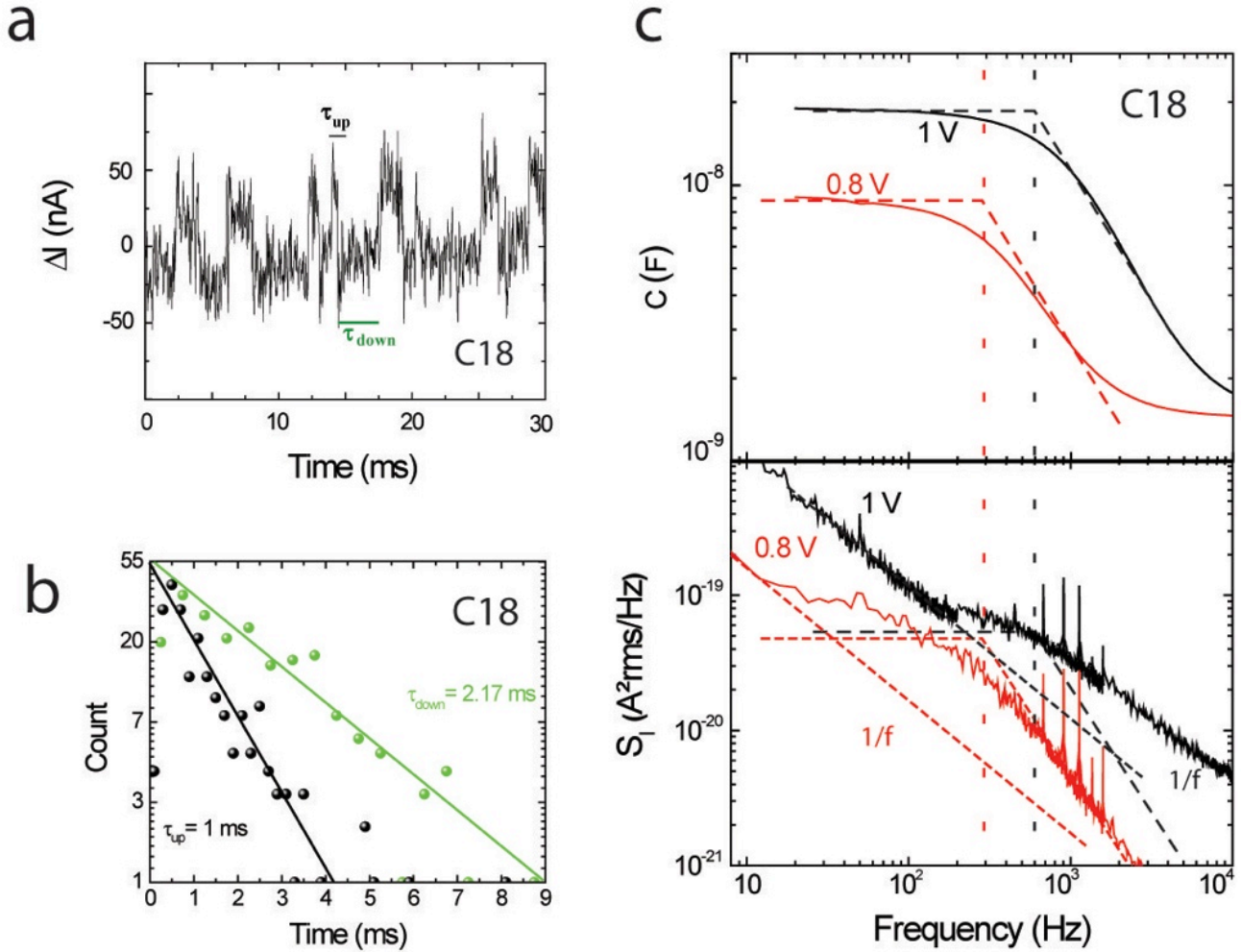

**Figure 6.** a) Random Telegraph Signal measured at 0.9 V for the as-fabricated samples C. $\Delta I$-$t$ is the time-dependent fluctuation of current around the averaged current value $I$=5.5 $\mu$A.

b) Histograms on the upper $\tau_{up}$ and lower $\tau_{down}$ characteristic times (as defined Fig. 6-a) of RTS noise measured on sample C.

c) $C$-$f$ (capacitance-frequency) and power spectrum current noise $S_I$ measured at resonance $V_{RES}$ = 0.9 V. $S_I$ is composed of $1/f$ noise and Lorentzian spectrum due to presence of RTS. Corner frequency, $f_c$, obtained from *C-f* is indicated as dashed lines and is in good agreement with $f_{cRTS}$, the corner frequency of the RTS Lorentzian noise.

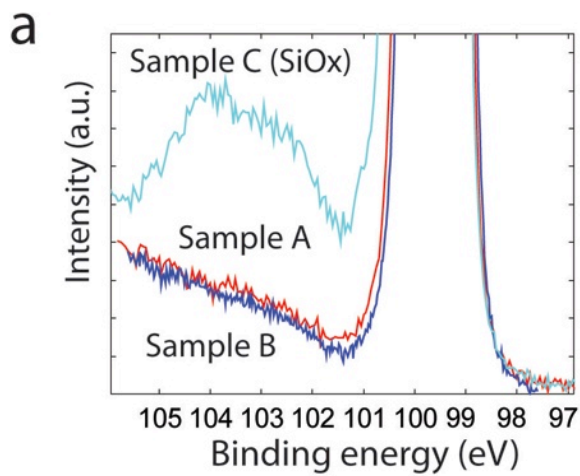
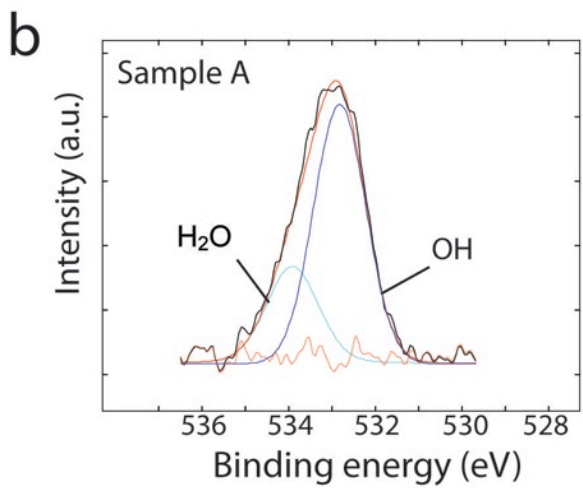
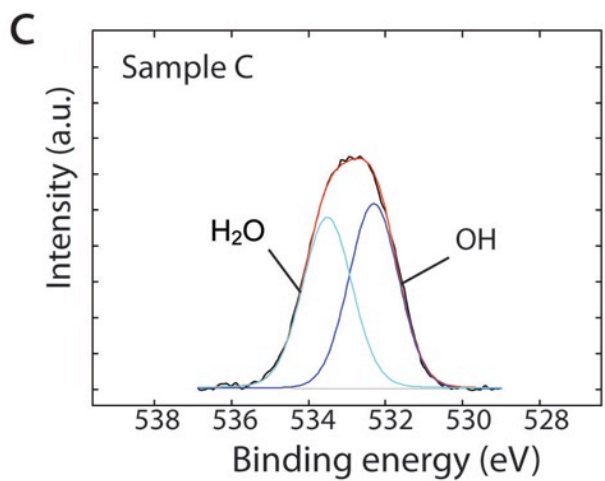

**Figure 7.** a) XPS Si2p spectra for freshly prepared samples A (red), B (blue) and C (cyan). b) and c) O1s core level peak decomposition for sample A and C respectively.

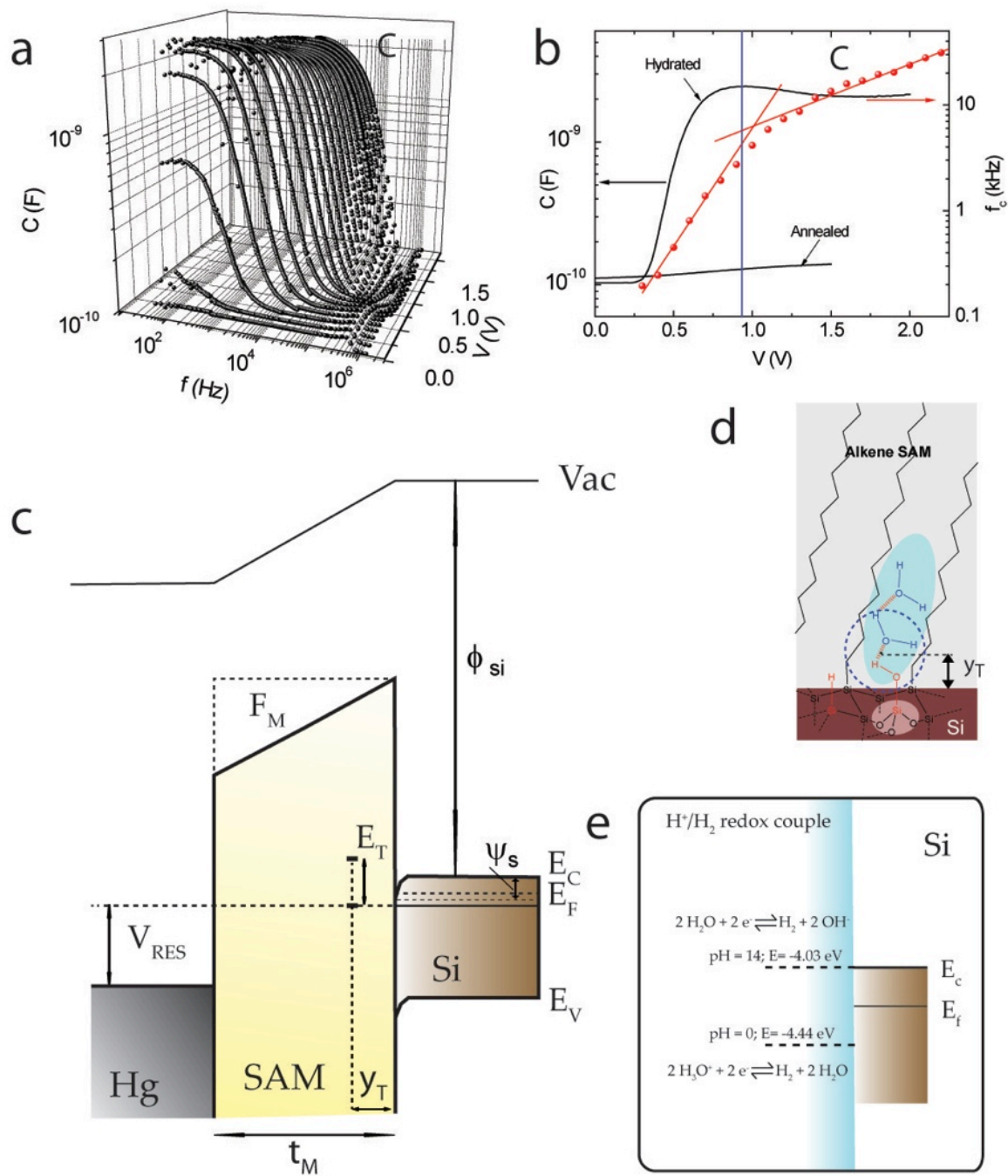

**Figure 8.** a) Capacitance *C* plotted as a function of voltage and frequency for sample C.

b) Capacitance $C$ at 1 kHz (for both annealed and hydrated samples) and corner frequency $f_c$ (extracted from data in 7-a following the definition shown in Fig. 5-a), plotted as a function of $V$ for sample C.

c) Schematic energy band diagram of the molecular junction with introduction of an energy level $E_T$ in the HOMO-LUMO band gap of the monolayer at a distance $y_T$ from the Si surface. $F_T$ is the electric field across the SAM (considered constant), $t_M$ the SAM thickness (Table 1). $V_{ac}$ is the vacuum energy level, $\psi_s$ the silicon surface potential, $E_C$, $E_F$ and $E_V$ are silicon conduction band, Fermi and Valence band energy levels, respectively. At $V = V_{RES}$, the trap energy level is shifted down by $E_T$ and is aligned with the Fermi energy of the silicon substrate.

d) Schematic view of alkene monolayers on silicon substrate with adsorption of water molecules at the bottom part of the monolayers. We consider $y_T$ the distance between the silicon substrate and half of a water monolayer thickness (2.5 Å from ref. 42)

e) Energy level diagram of the $H^+/H_2$ redox couple (left) compared to the silicon band structure at flat-band condition ($V \sim 0V$). The two reversible reactions involving $H_2O$ are shown. The trap energy level is related to the pH : -4.44 eV when pH = 0 and -4.03 eV when pH=-4.44 eV.[45]

TABLES.

**Table 1.** Summary of the key parameters for samples A, B and C. O1s-$H_2O$ is the area of the $H_2O$ contribution in O1s peak. $t_M$ is the measured thickness of the SAM, C.A. is the water contact angle.

| | Resistivity (Ω.cm) | Reaction time | $t_M$ (Å) | C.A. (°) | [O1s]/[Si2p+C1s] | O1s-$H_2O$/[Si2p+C1s] |
|---|---|---|---|---|---|---|
| Sample A | 1 | 30 mn | 17±1 | 108±1 | 0.015 | 0.003 |
| Sample B | 1 | 5 h | 20±1 | 110±1 | 0.009 | 0.002 |
| Sample C | 0.01 | 5 h | 19±1 | 109±1 | 0.091 | 0.01 |

**Table 2.** Summary of equivalent model (inset Fig. 5-a) parameters extracted from fitting eq. 1 on capacitance and conductance data for the hydrated sample shown in Fig. 5.

| $G_{SAM}$ (S) | $G_{H2O}$ (S) | $\tau$ (s) | $C_{SAM}$ (F) | $C_{H2O}$ (F) | tg δ |
|---|---|---|---|---|---|
| $4 \times 10^{-7}$ | $2.2 \times 10^{-5}$ | $1.4 \times 10^{-4}$ | $1.6 \times 10^{-10}$ | $6 \times 10^{-10}$ | 0.01 |

SYNOPSIS TOC. Electronic transport through molecular junctions on silicon strongly depends on the presence of water between molecules. Signatures are increase of conductance by up to three orders of magnitude, asymmetry in *I-V* curves, increase of capacitance at low-frequency and anomalous Random-Telegraph-Signal.

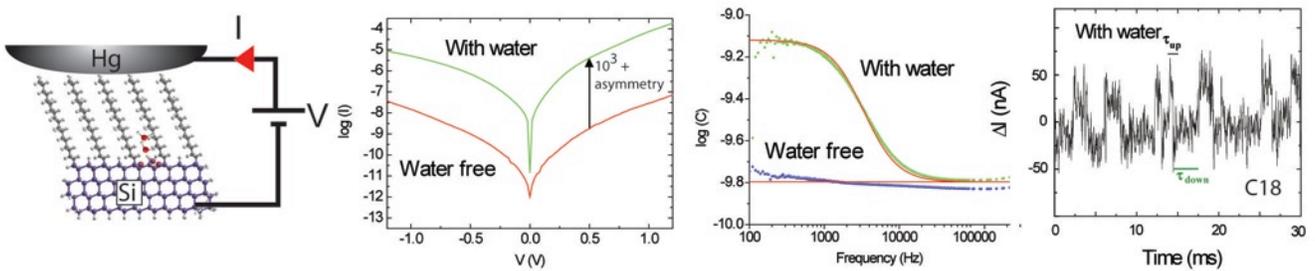

# Role of Hydration on the Electronic Transport through Molecular Junctions on Silicon

*Nicolas Clément, David Guérin, Stéphane Pleutin, Sylvie Godey and Dominique Vuillaume*

IEMN-CNRS, avenue Poincaré, Cité scientifique, 59652 Villeneuve d'Acq, France

# Supporting information

## 1) *I-V* curves and *C-V* curves samples A and B

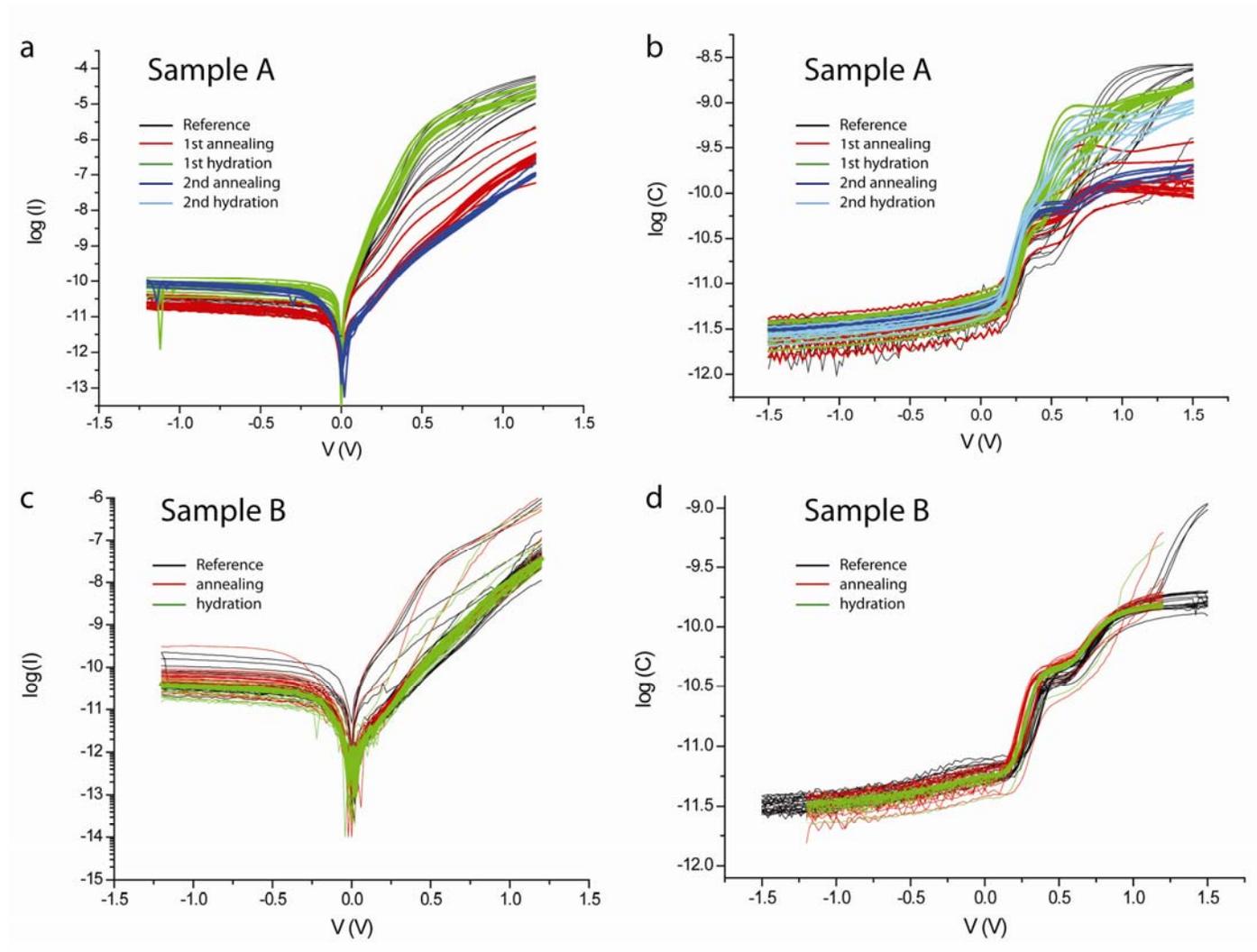

**Figure S1.** a) *I-V* (current-voltage) for sample A. b) *C@1kHz-V* (capacitance-voltage) for sample A. c) *I-V* for sample B. d) *C@1 kHz-V* (capacitance-voltage) for sample B.

## 2) *I-V* curves for sample C

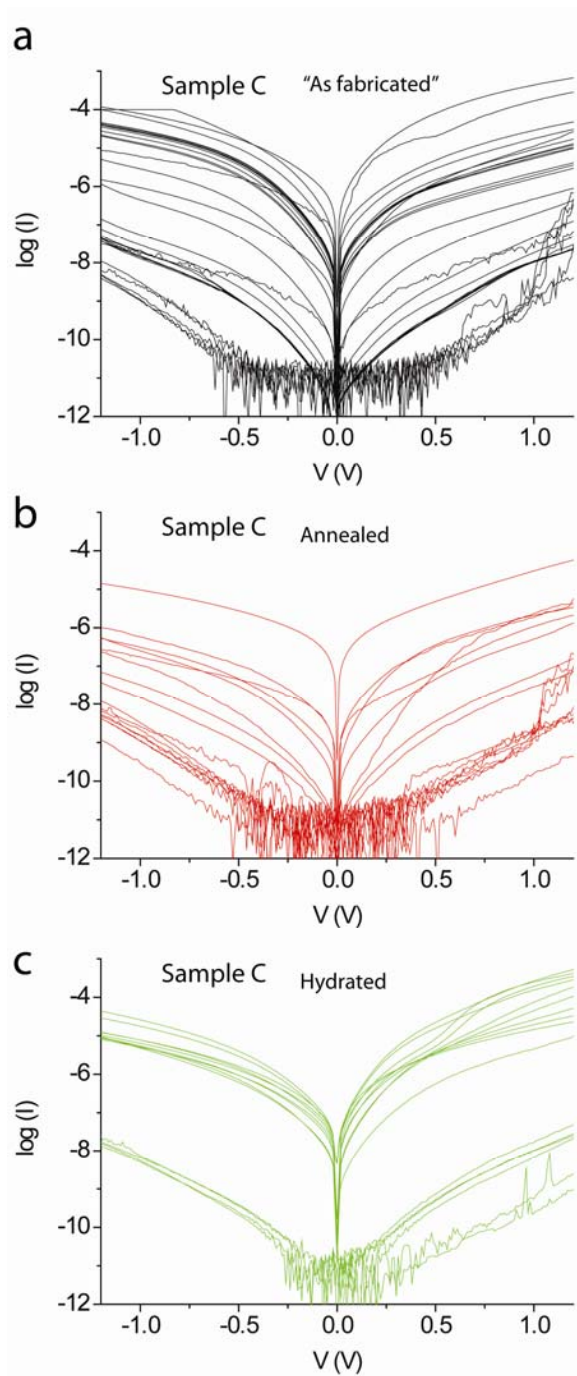

**Figure S2.** *I-V* (current-voltage) for sample C: a) "as fabricated", b) annealed, c) hydrated samples.

**3) Sample C, C14**

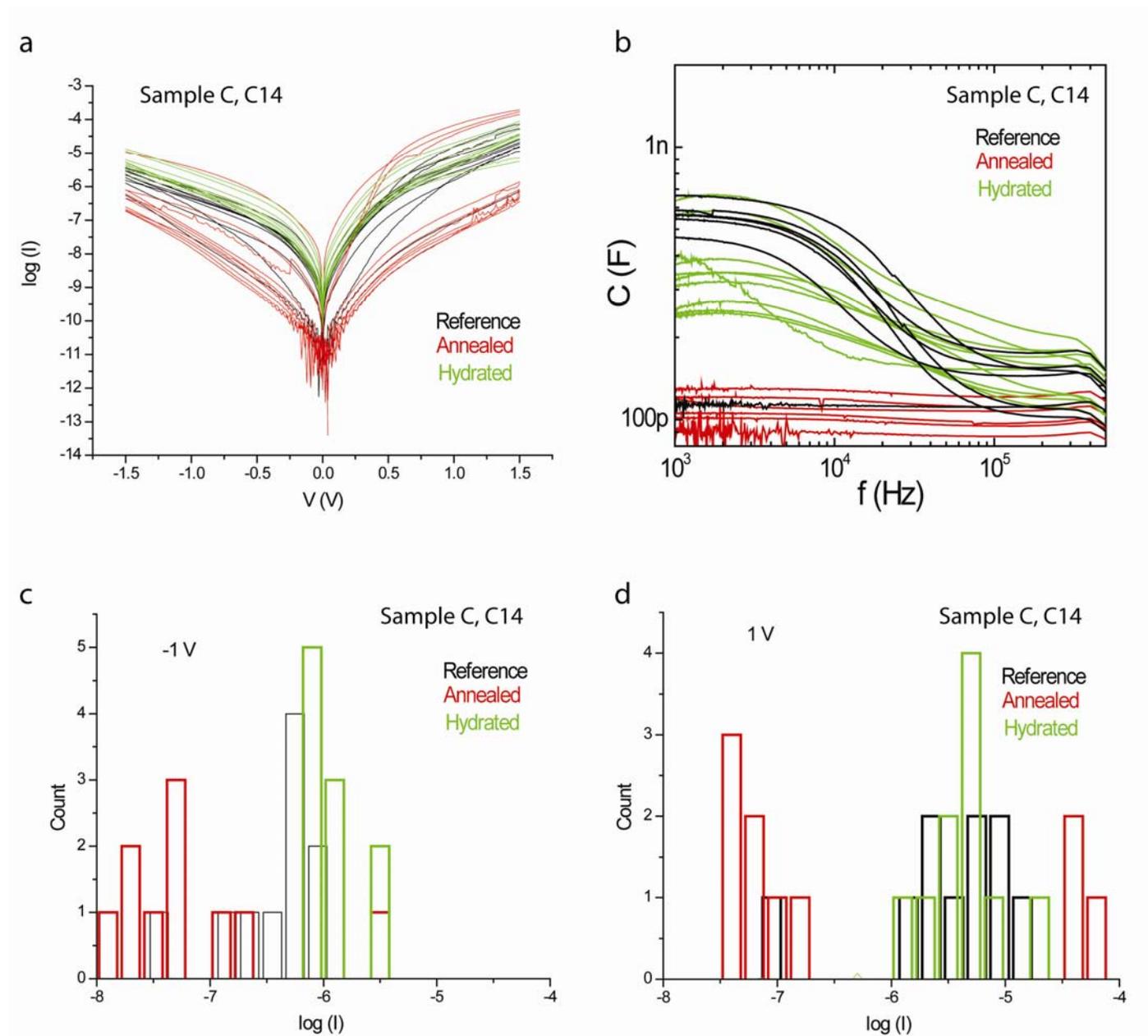

**Figure S3** a) *I-V* for sample A. b) *C-f* for sample A. c) Histograms of *I* at -1 V. d) Histograms of *I* at +1 V.

## 4) Low-frequency noise measurements for annealed sample B

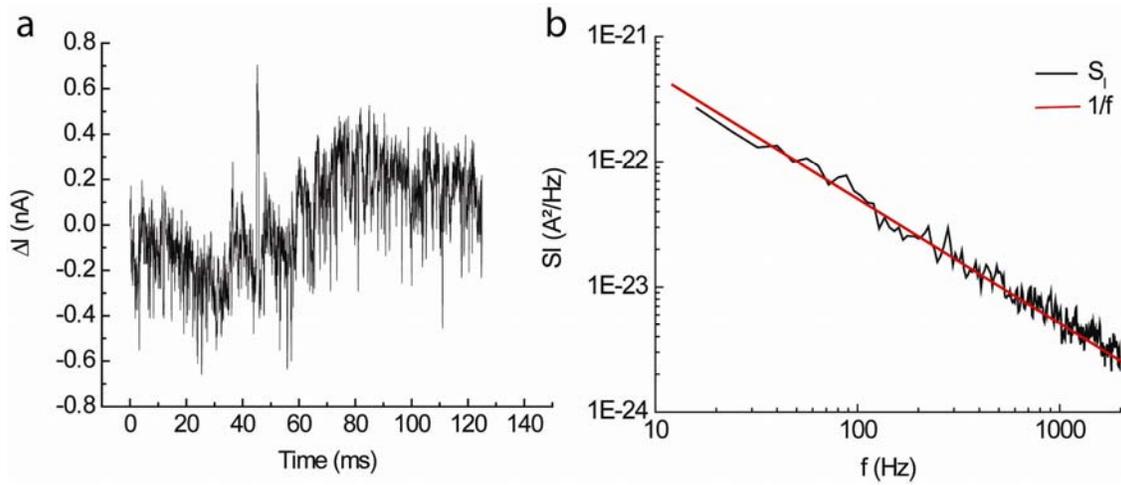

**Figure S4** a) I-t measurement at *V=0.8* V. No RTS is observed. b) Power spectrum current noise at *V=0.8* V. It follows exactly *1/f* dependence with no Lorentzian contribution (as it would has been expected with RTS).

## 5) Estimation of $\psi_s$ for samples A and C

From Ref. S1, we write $\psi_s = -\phi_T \ln\left[\dfrac{\left(\dfrac{V-V_{FB}}{k_0}\right)^2 + \phi_T}{\phi_T}\right]$, where $\phi_T = kT/q$, k the Boltzmann factor, T the temperature, q the electron charge, $V_{FB}$ the flat-band voltage (~ 0 V) and $k_0$ is the body effect factor for bulk with $k_0 = (2.q.\varepsilon_{Si}.N_D)^{0.5}/C_{Si}$. $\varepsilon_{Si}$ is the dielectric constant of silicon and $N_D \approx 10^{18}$ cm$^{-3}$ and $5\times10^{15}$ cm$^{-3}$ the doping level of bulk silicon for sample C and A respectively and $C_{Si}$ the silicon capacitance. Fig.S3 shows the simulation of $\psi_S$-V.

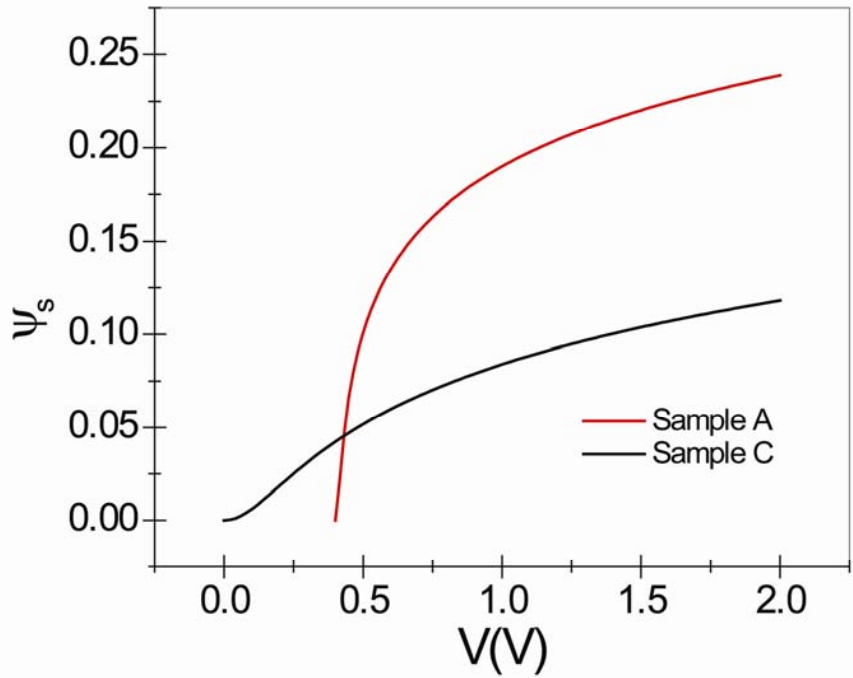

**Figure S5.** $\psi_S$-V for samples A and C.

## 6) $V_{RES}$ evaluation for sample A

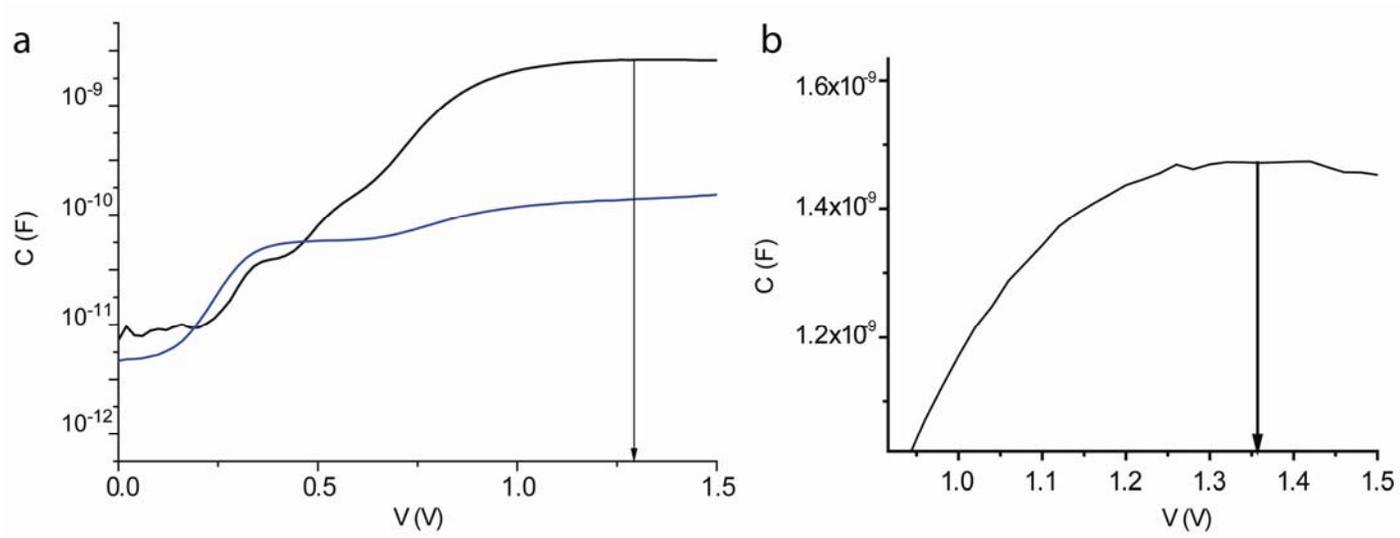

**Figure S6.** a) *C-V* measurements on sample A (black: hydrated and blue annealed). b) Zoom of a) for hydrated sample A in linear sacle. $V_{RES}$ is indicated by an arrow on a) and b)

## 7) *C-f* measurements for C18-thiol molecular junctions

We prepared gold Au(111) substrates by evaporating 2 nm of titanium to promote adhesion and 200 nm of gold onto cleaned silicon substrate using an e-beam evaporator. We used a low deposition rate (3 Å . s$^{-1}$) at 10$^{-8}$ Torr to minimize the roughness. For the SAM fabrication, we exposed this freshly prepared gold surface to 1 mM solution of $C_{18}H_{37}SH$ molecules in dichloromethane during 24 h. Then, we rinsed the treated substrates with dichloromethane followed by a cleaning in an ultrasonic bath of dichloromethane during 5 min.

For electrical measurements, we couldn't use Hg as an upper electrode due to its affinity with gold. We used Eutectic GaIn drop contact (eGaIn 99.99%, Ga:In; 75.5:24.5 wt % from Alfa Aesar) instead. We used a method close to the one developed by Chiechi et al.[S2] We formed a drop of eGaIn at the extremity of a needle fixed on a micromanipulator. By displacing the needle, we brought the drop into contact with a sacrificial surface, and we retracted the needle slowly. By this technique, we formed a conical tip of eGaIn with a diameter ranging from 50 to 200 μm (corresponding to contact area ranging from 10$^{-5}$ to 10$^{-3}$ cm$^2$). This conical tip was then put into contact with SAM (under control with a digital video camera). Voltage was applied on the eGaIn drop.

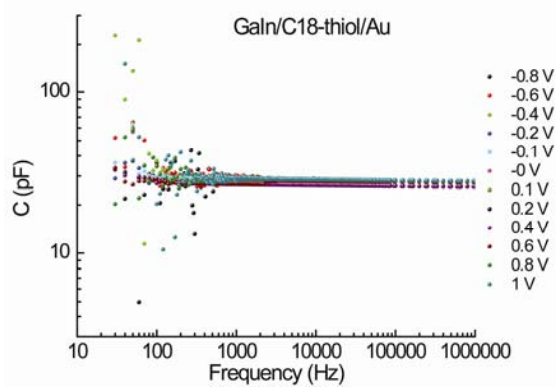

**Figure S7.** *C-f* measurements on GaIn/C18-thiol/Au molecular junctions with *V* from -0.8 V to 1 V by 0.2 V steps. No clear increases of capacitance are observed at low frequency as for molecular junctions on silicon.